\documentstyle[preprint,aps,prb]{revtex}
\tightenlines

\def\PsfigVersion{1.9}
\ifx\undefined\psfig\else \fi

\let\LaTeXAtSign=\@
\let\@=\relax
\edef\psfigRestoreAt{\catcode`\@=\number\catcode`@\relax}
\catcode`\@=11\relax
\newwrite\@unused
\def\ps@typeout#1{{\let\protect\string\immediate\write\@unused{#1}}}
\ps@typeout{psfig/tex \PsfigVersion}

\def\figurepath{./}

\def\@nnil{\@nil}
\def\@empty{}
\def\@psdonoop#1\@@#2#3{}
\def\@psdo#1:=#2\do#3{\edef\@psdotmp{#2}\ifx\@psdotmp\@empty \else
    \expandafter\@psdoloop#2,\@nil,\@nil\@@#1{#3}\fi}
\def\@psdoloop#1,#2,#3\@@#4#5{\def#4{#1}\ifx #4\@nnil \else
       #5\def#4{#2}\ifx #4\@nnil \else#5\@ipsdoloop #3\@@#4{#5}\fi\fi}
\def\@ipsdoloop#1,#2\@@#3#4{\def#3{#1}\ifx #3\@nnil 
       \let\@nextwhile=\@psdonoop \else
      #4\relax\let\@nextwhile=\@ipsdoloop\fi\@nextwhile#2\@@#3{#4}}
\def\@tpsdo#1:=#2\do#3{\xdef\@psdotmp{#2}\ifx\@psdotmp\@empty \else
    \@tpsdoloop#2\@nil\@nil\@@#1{#3}\fi}
\def\@tpsdoloop#1#2\@@#3#4{\def#3{#1}\ifx #3\@nnil 
       \let\@nextwhile=\@psdonoop \else
      #4\relax\let\@nextwhile=\@tpsdoloop\fi\@nextwhile#2\@@#3{#4}}
\ifx\undefined\fbox
\newdimen\fboxrule
\newdimen\fboxsep
\newdimen\ps@tempdima
\newbox\ps@tempboxa
\long\def\fbox#1{\leavevmode\setbox\ps@tempboxa\hbox{#1}\ps@tempdima\fboxrule
    \advance\ps@tempdima \fboxsep \advance\ps@tempdima \dp\ps@tempboxa
   \hbox{\lower \ps@tempdima\hbox
  {\vbox{\hrule height \fboxrule
          \hbox{\vrule width \fboxrule \hskip\fboxsep
          \vbox{\vskip\fboxsep \box\ps@tempboxa\vskip\fboxsep}\hskip 
                 \fboxsep\vrule width \fboxrule}
                 \hrule height \fboxrule}}}}
\fi
\newread\ps@stream
\newif\ifnot@eof       
\newif\if@noisy        
\newif\if@atend        
\newif\if@psfile       
{\catcode`\%=12\global\gdef\epsf@start{
\def\epsf@PS{PS}
\def\epsf@getbb#1{%
\openin\ps@stream=#1
\ifeof\ps@stream\ps@typeout{Error, File #1 not found}\else
   {\not@eoftrue \chardef\other=12
    \def\do##1{\catcode`##1=\other}\dospecials \catcode`\ =10
    \loop
       \if@psfile
	  \read\ps@stream to \epsf@fileline
       \else{
	  \obeyspaces
          \read\ps@stream to \epsf@tmp\global\let\epsf@fileline\epsf@tmp}
       \fi
       \ifeof\ps@stream\not@eoffalse\else
       \if@psfile\else
       \expandafter\epsf@test\epsf@fileline:. \\%
       \fi
          \expandafter\epsf@aux\epsf@fileline:. \\%
       \fi
   \ifnot@eof\repeat
   }\closein\ps@stream\fi}%
\long\def\epsf@test#1#2#3:#4\\{\def\epsf@testit{#1#2}
			\ifx\epsf@testit\epsf@start\else
\ps@typeout{Warning! File does not start with `\epsf@start'.  It may not be a PostScript file.}
			\fi
			\@psfiletrue} 
{\catcode`\%=12\global\let\epsf@percent=
\long\def\epsf@aux#1#2:#3\\{\ifx#1\epsf@percent
   \def\epsf@testit{#2}\ifx\epsf@testit\epsf@bblit
	\@atendfalse
        \epsf@atend #3 . \\%
	\if@atend	
	   \if@verbose{
		\ps@typeout{psfig: found `(atend)'; continuing search}
	   }\fi
        \else
        \epsf@grab #3 . . . \\%
        \not@eoffalse
        \global\no@bbfalse
        \fi
   \fi\fi}%
\def\epsf@grab #1 #2 #3 #4 #5\\{%
   \global\def\epsf@llx{#1}\ifx\epsf@llx\empty
      \epsf@grab #2 #3 #4 #5 .\\\else
   \global\def\epsf@lly{#2}%
   \global\def\epsf@urx{#3}\global\def\epsf@ury{#4}\fi}%
\def\epsf@atendlit{(atend)} 
\def\epsf@atend #1 #2 #3\\{%
   \def\epsf@tmp{#1}\ifx\epsf@tmp\empty
      \epsf@atend #2 #3 .\\\else
   \ifx\epsf@tmp\epsf@atendlit\@atendtrue\fi\fi}

\chardef\psletter = 11 
\chardef\other = 12

\newif \ifdebug 
\newif\ifc@mpute 
\c@mputetrue 

\let\then = \relax
\def\r@dian{pt }
\let\r@dians = \r@dian
\let\dimensionless@nit = \r@dian
\let\dimensionless@nits = \dimensionless@nit
\def\internal@nit{sp }
\let\internal@nits = \internal@nit
\newif\ifstillc@nverging
\def \Mess@ge #1{\ifdebug \then \message {#1} \fi}

{ 
	\catcode `\@ = \psletter
	\gdef \nodimen {\expandafter \n@dimen \the \dimen}
	\gdef \term #1 #2 #3%
	       {\edef \t@ {\the #1}
		\edef \t@@ {\expandafter \n@dimen \the #2\r@dian}%
		\t@rm {\t@} {\t@@} {#3}%
	       }
	\gdef \t@rm #1 #2 #3%
	       {{%
		\count 0 = 0
		\dimen 0 = 1 \dimensionless@nit
		\dimen 2 = #2\relax
		\Mess@ge {Calculating term #1 of \nodimen 2}%
		\loop
		\ifnum	\count 0 < #1
		\then	\advance \count 0 by 1
			\Mess@ge {Iteration \the \count 0 \space}%
			\Multiply \dimen 0 by {\dimen 2}%
			\Mess@ge {After multiplication, term = \nodimen 0}%
			\Divide \dimen 0 by {\count 0}%
			\Mess@ge {After division, term = \nodimen 0}%
		\repeat
		\Mess@ge {Final value for term #1 of 
				\nodimen 2 \space is \nodimen 0}%
		\xdef \Term {#3 = \nodimen 0 \r@dians}%
		\aftergroup \Term
	       }}
	\catcode `\p = \other
	\catcode `\t = \other
	\gdef \n@dimen #1pt{#1} 
}

\def \Divide #1by #2{\divide #1 by #2} 

\def \Multiply #1by #2
       {{
	\count 0 = #1\relax
	\count 2 = #2\relax
	\count 4 = 65536
	\Mess@ge {Before scaling, count 0 = \the \count 0 \space and
			count 2 = \the \count 2}%
	\ifnum	\count 0 > 32767 
	\then	\divide \count 0 by 4
		\divide \count 4 by 4
	\else	\ifnum	\count 0 < -32767
		\then	\divide \count 0 by 4
			\divide \count 4 by 4
		\else
		\fi
	\fi
	\ifnum	\count 2 > 32767 
	\then	\divide \count 2 by 4
		\divide \count 4 by 4
	\else	\ifnum	\count 2 < -32767
		\then	\divide \count 2 by 4
			\divide \count 4 by 4
		\else
		\fi
	\fi
	\multiply \count 0 by \count 2
	\divide \count 0 by \count 4
	\xdef \product {#1 = \the \count 0 \internal@nits}%
	\aftergroup \product
       }}

\def\r@duce{\ifdim\dimen0 > 90\r@dian \then   
		\multiply\dimen0 by -1
		\advance\dimen0 by 180\r@dian
		\r@duce
	    \else \ifdim\dimen0 < -90\r@dian \then  
		\advance\dimen0 by 360\r@dian
		\r@duce
		\fi
	    \fi}

\def\Sine#1%
       {{%
	\dimen 0 = #1 \r@dian
	\r@duce
	\ifdim\dimen0 = -90\r@dian \then
	   \dimen4 = -1\r@dian
	   \c@mputefalse
	\fi
	\ifdim\dimen0 = 90\r@dian \then
	   \dimen4 = 1\r@dian
	   \c@mputefalse
	\fi
	\ifdim\dimen0 = 0\r@dian \then
	   \dimen4 = 0\r@dian
	   \c@mputefalse
	\fi
	\ifc@mpute \then
		\divide\dimen0 by 180
		\dimen0=3.141592654\dimen0
		\dimen 2 = 3.1415926535897963\r@dian 
		\divide\dimen 2 by 2 
		\Mess@ge {Sin: calculating Sin of \nodimen 0}%
		\count 0 = 1 
		\dimen 2 = 1 \r@dian 
		\dimen 4 = 0 \r@dian 
		\loop
			\ifnum	\dimen 2 = 0 
			\then	\stillc@nvergingfalse 
			\else	\stillc@nvergingtrue
			\fi
			\ifstillc@nverging 
			\then	\term {\count 0} {\dimen 0} {\dimen 2}%
				\advance \count 0 by 2
				\count 2 = \count 0
				\divide \count 2 by 2
				\ifodd	\count 2 
				\then	\advance \dimen 4 by \dimen 2
				\else	\advance \dimen 4 by -\dimen 2
				\fi
		\repeat
	\fi		
			\xdef \sine {\nodimen 4}%
       }}

\def\Cosine#1{\ifx\sine\UnDefined\edef\Savesine{\relax}\else
		             \edef\Savesine{\sine}\fi
	{\dimen0=#1\r@dian\advance\dimen0 by 90\r@dian
	 \Sine{\nodimen 0}
	 \xdef\cosine{\sine}
	 \xdef\sine{\Savesine}}}	      

\def\psdraft{
	\def\@psdraft{0}
}
\def\psfull{
	\def\@psdraft{100}
}

\psfull

\newif\if@scalefirst
\def\psscalefirst{\@scalefirsttrue}
\def\psrotatefirst{\@scalefirstfalse}
\psrotatefirst

\newif\if@draftbox
\def\psnodraftbox{
	\@draftboxfalse
}
\def\psdraftbox{
	\@draftboxtrue
}
\@draftboxtrue

\newif\if@prologfile
\newif\if@postlogfile
\def\pssilent{
	\@noisyfalse
}
\def\psnoisy{
	\@noisytrue
}
\psnoisy
\newif\if@bbllx
\newif\if@bblly
\newif\if@bburx
\newif\if@bbury
\newif\if@height
\newif\if@width
\newif\if@rheight
\newif\if@rwidth
\newif\if@angle
\newif\if@clip
\newif\if@verbose
\newif\if@scale
\def\@p@@sclip#1{\@cliptrue}

\newif\if@decmpr

\def\@p@@sfigure#1{\def\@p@sfile{null}\def\@p@sbbfile{null}
	        \openin1=#1.bb
		\ifeof1\closein1
	        	\openin1=\figurepath#1.bb
			\ifeof1\closein1
			        \openin1=#1
				\ifeof1\closein1%
				       \openin1=\figurepath#1
					\ifeof1
					   \ps@typeout{Error, File #1 not found}
						\if@bbllx\if@bblly
				   		\if@bburx\if@bbury
			      				\def\@p@sfile{#1}%
			      				\def\@p@sbbfile{#1}%
							\@decmprfalse
				  	   	\fi\fi\fi\fi
					\else\closein1
				    		\def\@p@sfile{\figurepath#1}%
				    		\def\@p@sbbfile{\figurepath#1}%
						\@decmprfalse
	                       		\fi%
			 	\else\closein1%
					\def\@p@sfile{#1}
					\def\@p@sbbfile{#1}
					\@decmprfalse
			 	\fi
			\else
				\def\@p@sfile{\figurepath#1}
				\def\@p@sbbfile{\figurepath#1.bb}
				\@decmprtrue
			\fi
		\else
			\def\@p@sfile{#1}
			\def\@p@sbbfile{#1.bb}
			\@decmprtrue
		\fi}

\def\@p@@sfile#1{\@p@@sfigure{#1}}

\def\@p@@sbbllx#1{
		\@bbllxtrue
		\dimen100=#1
		\edef\@p@sbbllx{\number\dimen100}
}
\def\@p@@sbblly#1{
		\@bbllytrue
		\dimen100=#1
		\edef\@p@sbblly{\number\dimen100}
}
\def\@p@@sbburx#1{
		\@bburxtrue
		\dimen100=#1
		\edef\@p@sbburx{\number\dimen100}
}
\def\@p@@sbbury#1{
		\@bburytrue
		\dimen100=#1
		\edef\@p@sbbury{\number\dimen100}
}
\def\@p@@sheight#1{
		\@heighttrue
		\dimen100=#1
   		\edef\@p@sheight{\number\dimen100}
}
\def\@p@@swidth#1{
		\@widthtrue
		\dimen100=#1
		\edef\@p@swidth{\number\dimen100}
}
\def\@p@@srheight#1{
		\@rheighttrue
		\dimen100=#1
		\edef\@p@srheight{\number\dimen100}
}
\def\@p@@srwidth#1{
		\@rwidthtrue
		\dimen100=#1
		\edef\@p@srwidth{\number\dimen100}
}
\def\@p@@sangle#1{
		\@angletrue
		\edef\@p@sangle{#1} 
}
\def\@p@@srotate#1{\@p@@sangle{-#1}}
\def\@p@@sscale#1{
		\@scaletrue
		\edef\@p@sscale{#1}
}
\def\@p@@ssilent#1{ 
		\@verbosefalse
}
\def\@p@@sprolog#1{\@prologfiletrue\def\@prologfileval{#1}}
\def\@p@@spostlog#1{\@postlogfiletrue\def\@postlogfileval{#1}}
\def\@cs@name#1{\csname #1\endcsname}
\def\@setparms#1=#2,{\@cs@name{@p@@s#1}{#2}}
\def\ps@init@parms{
		\@bbllxfalse \@bbllyfalse
		\@bburxfalse \@bburyfalse
		\@heightfalse \@widthfalse
		\@rheightfalse \@rwidthfalse
		\@scalefalse
		\def\@p@sbbllx{}\def\@p@sbblly{}
		\def\@p@sbburx{}\def\@p@sbbury{}
		\def\@p@sheight{}\def\@p@swidth{}
		\def\@p@srheight{}\def\@p@srwidth{}
		\def\@p@sangle{0}
		\def\@p@sfile{} \def\@p@sbbfile{}
		\def\@p@scost{10}
		\def\@sc{}
		\@prologfilefalse
		\@postlogfilefalse
		\@clipfalse
		\if@noisy
			\@verbosetrue
		\else
			\@verbosefalse
		\fi
}
\def\parse@ps@parms#1{
	 	\@psdo\@psfiga:=#1\do
		   {\expandafter\@setparms\@psfiga,}}
\newif\ifno@bb
\def\bb@missing{
	\if@verbose{
		\ps@typeout{psfig: searching \@p@sbbfile \space  for bounding box}
	}\fi
	\no@bbtrue
	\epsf@getbb{\@p@sbbfile}
        \ifno@bb \else \bb@cull\epsf@llx\epsf@lly\epsf@urx\epsf@ury\fi
}	
\def\bb@cull#1#2#3#4{
	\dimen100=#1 bp\edef\@p@sbbllx{\number\dimen100}
	\dimen100=#2 bp\edef\@p@sbblly{\number\dimen100}
	\dimen100=#3 bp\edef\@p@sbburx{\number\dimen100}
	\dimen100=#4 bp\edef\@p@sbbury{\number\dimen100}
	\no@bbfalse
}
\newdimen\p@intvaluex
\newdimen\p@intvaluey
\def\rotate@#1#2{{\dimen0=#1 sp\dimen1=#2 sp
		  \global\p@intvaluex=\cosine\dimen0
		  \dimen3=\sine\dimen1
		  \global\advance\p@intvaluex by -\dimen3
		  \global\p@intvaluey=\sine\dimen0
		  \dimen3=\cosine\dimen1
		  \global\advance\p@intvaluey by \dimen3
		  }}
\def\compute@bb{
		\no@bbfalse
		\if@bbllx \else \no@bbtrue \fi
		\if@bblly \else \no@bbtrue \fi
		\if@bburx \else \no@bbtrue \fi
		\if@bbury \else \no@bbtrue \fi
		\ifno@bb \bb@missing \fi
		\ifno@bb \ps@typeout{FATAL ERROR: no bb supplied or found}
			\no-bb-error
		\fi
		\count203=\@p@sbburx
		\count204=\@p@sbbury
		\advance\count203 by -\@p@sbbllx
		\advance\count204 by -\@p@sbblly
		\edef\ps@bbw{\number\count203}
		\edef\ps@bbh{\number\count204}
		\if@angle 
			\Sine{\@p@sangle}\Cosine{\@p@sangle}
	        	{\dimen100=\maxdimen\xdef\r@p@sbbllx{\number\dimen100}
					    \xdef\r@p@sbblly{\number\dimen100}
			                    \xdef\r@p@sbburx{-\number\dimen100}
					    \xdef\r@p@sbbury{-\number\dimen100}}
                        \def\minmaxtest{
			   \ifnum\number\p@intvaluex<\r@p@sbbllx
			      \xdef\r@p@sbbllx{\number\p@intvaluex}\fi
			   \ifnum\number\p@intvaluex>\r@p@sbburx
			      \xdef\r@p@sbburx{\number\p@intvaluex}\fi
			   \ifnum\number\p@intvaluey<\r@p@sbblly
			      \xdef\r@p@sbblly{\number\p@intvaluey}\fi
			   \ifnum\number\p@intvaluey>\r@p@sbbury
			      \xdef\r@p@sbbury{\number\p@intvaluey}\fi
			   }
			\rotate@{\@p@sbbllx}{\@p@sbblly}
			\minmaxtest
			\rotate@{\@p@sbbllx}{\@p@sbbury}
			\minmaxtest
			\rotate@{\@p@sbburx}{\@p@sbblly}
			\minmaxtest
			\rotate@{\@p@sbburx}{\@p@sbbury}
			\minmaxtest
			\edef\@p@sbbllx{\r@p@sbbllx}\edef\@p@sbblly{\r@p@sbblly}
			\edef\@p@sbburx{\r@p@sbburx}\edef\@p@sbbury{\r@p@sbbury}
		\fi
		\count203=\@p@sbburx
		\count204=\@p@sbbury
		\advance\count203 by -\@p@sbbllx
		\advance\count204 by -\@p@sbblly
		\edef\@bbw{\number\count203}
		\edef\@bbh{\number\count204}
}
\def\in@hundreds#1#2#3{\count240=#2 \count241=#3
		     \count100=\count240	
		     \divide\count100 by \count241
		     \count101=\count100
		     \multiply\count101 by \count241
		     \advance\count240 by -\count101
		     \multiply\count240 by 10
		     \count101=\count240	
		     \divide\count101 by \count241
		     \count102=\count101
		     \multiply\count102 by \count241
		     \advance\count240 by -\count102
		     \multiply\count240 by 10
		     \count102=\count240	
		     \divide\count102 by \count241
		     \count200=#1\count205=0
		     \count201=\count200
			\multiply\count201 by \count100
		 	\advance\count205 by \count201
		     \count201=\count200
			\divide\count201 by 10
			\multiply\count201 by \count101
			\advance\count205 by \count201
		     \count201=\count200
			\divide\count201 by 100
			\multiply\count201 by \count102
			\advance\count205 by \count201
		     \edef\@result{\number\count205}
}
\def\ps@scaleinhundreds#1{
		\in@hundreds{#1}{\@p@sscale}{100}
		\edef#1{\@result}
}
\def\compute@wfromh{
		\in@hundreds{\@p@sheight}{\@bbw}{\@bbh}
		\edef\@p@swidth{\@result}
}
\def\compute@hfromw{
	        \in@hundreds{\@p@swidth}{\@bbh}{\@bbw}
		\edef\@p@sheight{\@result}
}
\def\compute@handw{
		\if@height 
			\if@width
			\else
				\compute@wfromh
			\fi
		\else 
			\if@width
				\compute@hfromw
			\else
				\edef\@p@sheight{\@bbh}
				\edef\@p@swidth{\@bbw}
			\fi
		\fi
}
\def\compute@resv{
		\if@rheight \else \edef\@p@srheight{\@p@sheight} \fi
		\if@rwidth \else \edef\@p@srwidth{\@p@swidth} \fi
}
\def\compute@sizes{
	\compute@bb
	\if@scalefirst\if@angle
	\if@width
	   \in@hundreds{\@p@swidth}{\@bbw}{\ps@bbw}
	   \edef\@p@swidth{\@result}
	\fi
	\if@height
	   \in@hundreds{\@p@sheight}{\@bbh}{\ps@bbh}
	   \edef\@p@sheight{\@result}
	\fi
	\fi\fi
	\compute@handw
	\compute@resv
	\if@scale
	   \if@verbose
	      \ps@typeout{(scaling by \@p@sscale)}%
	   \fi
	   \ps@scaleinhundreds{\@p@swidth}%
	   \ps@scaleinhundreds{\@p@sheight}%
	   \ps@scaleinhundreds{\@p@srwidth}%
	   \ps@scaleinhundreds{\@p@srheight}%
	\fi
}

\def\psfig#1{\vbox {
	\ps@init@parms
	\parse@ps@parms{#1}
	\compute@sizes
	\ifnum\@p@scost<\@psdraft{
		\special{ps::[begin] 	\@p@swidth \space \@p@sheight \space
				\@p@sbbllx \space \@p@sbblly \space
				\@p@sbburx \space \@p@sbbury \space
				startTexFig \space }
		\if@angle
			\special {ps:: \@p@sangle \space rotate \space} 
		\fi
		\if@clip{
			\if@verbose{
				\ps@typeout{(clip)}
			}\fi
			\special{ps:: doclip \space }
		}\fi
		\if@prologfile
		    \special{ps: plotfile \@prologfileval \space } \fi
		\if@decmpr{
			\if@verbose{
				\ps@typeout{psfig: including \@p@sfile.Z \space }
			}\fi
			\special{ps: plotfile "`zcat \@p@sfile.Z" \space }
		}\else{
			\if@verbose{
				\ps@typeout{psfig: including \@p@sfile \space }
			}\fi
			\special{ps: plotfile \@p@sfile \space }
		}\fi
		\if@postlogfile
		    \special{ps: plotfile \@postlogfileval \space } \fi
		\special{ps::[end] endTexFig \space }
		\vbox to \@p@srheight true sp{
			\hbox to \@p@srwidth true sp{
				\hss
			}
		\vss
		}
	}\else{
		\if@draftbox{		
			\hbox{\frame{\vbox to \@p@srheight true sp{
			\vss
			\hbox to \@p@srwidth true sp{ \hss \@p@sfile \hss }
			\vss
			}}}
		}\else{
			\vbox to \@p@srheight true sp{
			\vss
			\hbox to \@p@srwidth true sp{\hss}
			\vss
			}
		}\fi

	}\fi
}}
\psfigRestoreAt
\let\@=\LaTeXAtSign

\begin{document}
\title{A Stabilization Mechanism of Zirconia Based on Oxygen Vacancies
       Only\footnote{Supported by the European Science Foundation, by the
       EPSRC for funding under Grants No. L66908 and No. L08380, and by
       the European Communities HCM Network ``Electronic Structure
       Calculations of Materials Properties and Processes for Industry
       and Basic Science'' under grant No. ERBFMRXCT980178}}
\author{Stefano Fabris,\footnote{Present address: Max-Planck-Institut
       f\"ur Metallforschung, Heisenbergstr. 3, D-70569 Stuttgart,
       Germany.}  Anthony T. Paxton and Michael W. Finnis}
       \address{Atomistic Simulation Group, Department of Pure and Applied
       Physics, Queen's University, \\ Belfast BT7 1NN, United Kingdom}
\date{\today}
       \maketitle

\begin{abstract}
The microscopic mechanism leading to stabilization of cubic and
tetragonal forms of zirconia (ZrO$_2$) is analyzed by means of a
self-consistent tight-binding model. Using this model, energies and
structures of zirconia containing different vacancy concentrations are
calculated, equivalent in concentration to the charge compensating
vacancies associated with dissolved yttria (Y$_2$O$_3$) in the tetragonal
and cubic phase fields (3.2 and 14.4\% mol respectively). The
model is shown to predict the large relaxations around an oxygen vacancy,
and the clustering of vacancies along the $\langle 111 \rangle$
directions, in good agreement with experiments and first principles
calculations. The vacancies alone are shown to explain the stabilization
of cubic zirconia, and the mechanism is analyzed.
\end{abstract}


\section{Introduction}

 Stabilized zirconias are materials of outstanding technological
 importance~\cite{Heuer81}. Pure zirconia (ZrO$_2$) is monoclinic ($m$) at
 zero temperature (space group $P2_1/c$,
 Ref.~\onlinecite{McCullough59,Smith65}); upon increase of the temperature
 (at zero pressure) the material transforms to tetragonal ($t$) and then to
 a cubic ($c$) fluorite structure (space groups $P4_2/nmc$ and {\it
 Fm}3{\it m},
 respectively).~\cite{Howard88,Teufer62,Aldebert85,Ackermann77} These phase
 transitions induce large volume changes and make the pure material
 unsuitable for applications. The addition of lower-valence oxides like
 CaO, MgO, or Y$_2$O$_3$ disfavour the $m$ phase, stabilizing more
 symmetric structures with cubic and tetragonal symmetry.~\cite{Subbarao81}
 On increasing dopant concentration, the material transforms to a
 tetragonal ($t^*$) form, called partially stabilized, and then to a cubic
 ($c^*$) one, called fully stabilized. The amount of doping needed for
 stabilization is quite substantial (no less than 8 \% mol Y$_2$O$_3$ to
 achieve full stabilization~\cite{Grain67,Stubican77,Goff99}), and the
 electrostatic neutrality of this rather ionic material is maintained by
 oxygen vacancies. In this case one vacancy is required for every two
 yttrium ions.

 The simultaneous presence of dopant cations and oxygen vacancies in large
 concentration means that the local atomic environments in the stabilized
 material are very different from the corresponding stoichiometric ($t$ and
 $c$) phases.  Despite the analogy in the sequences $m$--$t$--$c$ and
 $m$--$t^*$--$c^*$, there is no clear picture of the microscopic mechanisms
 of stabilization to parallel our understanding of the pure material: the
 most relevant issue concerns the respective roles of impurity cations and
 of oxygen vacancies~\cite{Stefanovich94}. 

 We address this issue by means of a self-consistent tight-binding model,
 which was parameterized on the electronic and structural properties of
 pure zirconia~\cite{Finnis98,strcZrO2}. The method was used to calculate
 the temperature evolution of the free energy surfaces governing the $c-t$
 phase transition, which was then described within the Landau theory of
 phase transformations~\cite{mdZrO2}.  These results are the starting point
 for the present approach to the stabilization mechanism, which is based on
 the qualitative analogy between isothermal and isoconcentration lines in
 the high-temperature region of the ZrO$_2$-Y$_2$O$_3$ phase-diagram. At
 fixed concentration, higher temperatures destabilize the tetragonal phases
 $t$ and $t^*$, favouring the cubic ones $c$ and $c^*$. Similarly, at fixed
 temperature, higher amount of impurities stabilize the cubic phases.

 The position of the oxygen vacancy which is associated with each pair of Y
 atoms has been a subject of some discussion.  Even though there are
 experimental data supporting the case in which Y is nearest neighbor (NN)
 to the vacancy~\cite{Tuilier87,Morikawa88,Steele74}, the most recent
 experiments~\cite{Catlow86,Veal88,Komyoji92,Li93a,Li93b,Li93c,Li94a,Li94b,Li94c,Goff99}
 suggest that Y is next nearest neighbor (NNN) to the vacancy.
 First-principles calculations~\cite{Stapper99,Bogicevic01} agree with the
 latter result, showing that the Y NNN position is energetically favoured
 by 0.34 eV with respect to the NN position.~\cite{Stapper99} The analysis
 of Li and coworkers~\cite{Li93a,Li93b,Li93c,Li94a,Li94b,Li94c} suggests
 that the Y atoms have a composition-independent 8 fold coordination shell
 (like in the perfect fluorite structure $c$) in both the $c^\ast$ and
 $t^\ast$ lattice types, further supporting the hypothesis that the dopants
 are on average NNN to the vacancies.  The presence of oxygen vacancies,
 together with the Y in the NNN position, reduces the average coordination
 number of the zirconium atoms from 8, as in the $c$ structure, to values
 closer to 7, similar to the $m$ phase.
 Besides the fact that the Y atoms are 8-fold coordinated in both the
 stabilized structures, on the basis of the $K$ edge in x-ray-absorption
 spectra, Li and coworkers~\cite{Li93a,Li93b,Li93c,Li94a,Li94b,Li94c}
 conclude that the perturbation in the neighborhood of the dopant cations
 is small and short-ranged.

 From these considerations, it is tempting to propose a physical picture in
 which the dopant cations, located in fluorite-like cation lattice sites
 NNN to the vacancies, do not take an active part in the stabilization
 mechanism, which is dominated by the crystal distortions around the oxygen
 vacancies. We now investigate the consequences of this hypothesis, by
 neglecting altogether the presence of the dopant atoms, therefore
 isolating the role played by oxygen vacancies in the stabilization of the
 $c^\ast$ and $t^\ast$ forms of zirconia.

 In Section~\ref{isolvac}, the model is shown to reproduce the local atomic
 and electronic structure around an isolated oxygen vacancy predicted by
 first-principles calculations. The atomistic structures resulting from
 modelling partially and fully stabilized zirconia are then presented in
 Secion~\ref{stab}, and the results are discussed in
 Section~\ref{disc}. Final remarks and a summary are made in
 Section~\ref{concl}.

\section{The isolated vacancy: structural and electronic properties}
\label{isolvac}

 As a preliminary step towards the modelling of stabilized zirconia, the
 structural and electronic properties of a model crystal structure
 containing an isolated oxygen vacancy are first addressed. The same system
 was studied from first principles by Stapper and
 coworkers:~\cite{Stapper99} a cubic 96-sites super-cell of 95 atoms in the
 fluorite structure containing one vacant oxygen lattice site, that we
 define as V$_1$. The vacancy is in the $+2$ charge state (one O$^{2-}$ ion
 missing). In the real material, the charge of the ${\rm V}_{\rm
 O}^{^{\cdot \cdot}}$ defect is compensated by the dopant substitutional
 cations ${\rm Y}_{\rm Zr}^\prime$. In our analysis, we mimic the presence
 of the charge-compensating Y atoms by distributing their charge over the
 32 Zr cations, so that the cell is electrostatically neutral. The
 electronic structure in the neighborhood of the vacancy is described by
 the dangling orbitals of its neighbouring atoms.

 We define as {\it unrelaxed} the configuration in which all the atoms are
 in the centro-symmetric fluorite positions, and as {\it relaxed} the
 crystal structure resulting from the static optimization of the atomic
 coordinates.

\begin{figure*}
\caption{Crystal relaxation of the atoms neighbouring an isolated
 vacancy in a fluorite lattice. Key: Zr black, O dark grey, vacancy light
 grey.}
\label{disp}
\begin{center}
\parbox{6cm}{\psfig{file=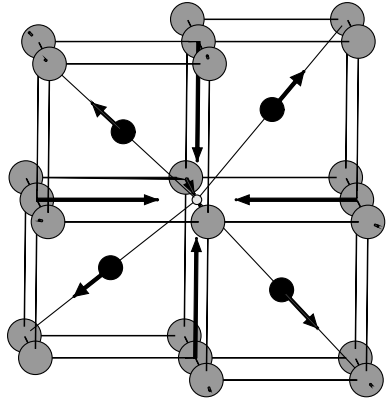,width=5cm,angle=-90}} 
\end{center}
\end{figure*}

 As the crystal is allowed to relax the Zr atoms NN to the vacancy radially
 shift outward along $\langle 111 \rangle$ directions, and the first shell
 of anions around the vacancy (NN O atoms) contract inward along $\langle
 100 \rangle$ directions. This relaxation pattern is driven by the
 electrostatics and is shown in Figure~\ref{disp}. The atomic displacements
 predicted by the SC-TB are compared to the first-principles
 results~\cite{Stapper99} in Table~\ref{tab}.  It may be noticed how the
 first-principles structural relaxations are well reproduced by the SC-TB
 model, {\it without} the inclusion of extra adjustable parameters.

\begin{figure}
\caption{Total density of states for the relaxed and unrelaxed
 configurations of the $V_1$ supercell, calculated with the SC-TB. The 0
 of energy is the highest occupied electron state. The arrow points to the
 energy levels of the Zr atoms NN to the vacancy, split from the $d$
 bands of the ``bulk'' Zr atoms.}
\label{dos}
\centerline{\psfig{file=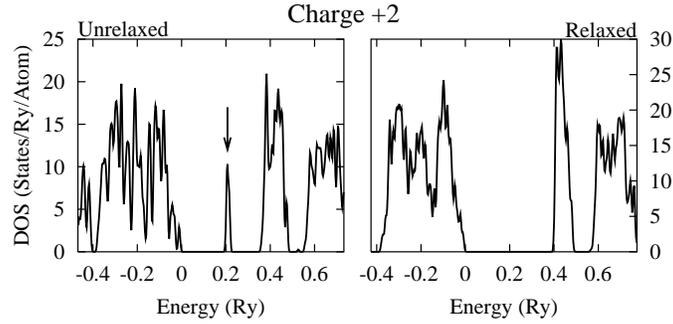,width=9cm,angle=0}} 
\end{figure}

 Figure~\ref{dos} shows the total density of states (DOS) calculated with
 the SC-TB model for both the relaxed and unrelaxed configurations.  The
 DOS are plotted with respect to the energy of the highest occupied
 level. Because of the choice of the reference energy, the valence 2$p$
 oxygen bands have negative energies. Above the bandgap, it is possible to
 identify the crystal field splitting of the zirconium $d$ bands. In the
 undistorted case (left panel), the arrow points to a sharp peak within the
 bandgap, also predicted by other calculations.~\cite{Kulkova93,Olkhovic95}
 The corresponding partial DOS shows that this peak corresponds to the NN
 Zr $d$ bands with symmetry $E$. The atomic relaxation (right panel) shifts
 the isolated peak back into the conduction band, as predicted by first
 principles calculations.~\cite{Stapper99}

 These results of the SC-TB model may be interpreted by considering the
 perturbation in the periodic electrostatic (Madelung) potential in the
 stoichiometric $c$ structure caused by the presence of an oxygen vacancy
 $V_O^{..}$. This defect increases the Madelung potential on the
 neighbouring sites, thereby lowering the energy of the $d$ states of the
 Zr ions NN to the vacancy with respect to the $d$ states of the bulk Zr
 ions far from the defect.  This perturbation in the Madelung potential at
 the nearest neighbour Zr sites is responsible for the isolated peak in the
 bandgap of the unrelaxed defective structure, marked by the arrow in
 Figure 2. The atomic relaxation effectively restores the homogeneity of
 the Madelung potential in the neighbourhood of the vacancy, pushing the NN
 Zr $d$ levels back upwards.

\section{Stabilised zirconia: structural stability}
\label{stab}

 The study of the stabilization mechanism is based on static and dynamic
 simulations of two 96-site super-cells which were chosen as representative
 systems. The first one, denoted as V$_1$, contains one vacancy. This
 system corresponds to 3.2\% mol Y$_2$O$_3$ and therefore it should be in
 the field of stability of the $t^\ast$ phase. Similarly, the second
 super-cell V$_4$ contains 4 vacancies and corresponds to 14.4\% mol
 Y$_2$O$_3$, which is in the phase-field of fully stabilized zirconia
 $c^\ast$. The electrostatic neutrality of the super-cells was ensured by
 distributing the compensating ionic charge of the dopant atoms over the 32
 Zr cations.  For each system, two separate sets of calculations were
 carried out: static relaxations and molecular dynamics (MD). The dynamics
 of the systems were simulated for 25 ps with a time-step of 5 fs,
 constraining the temperature at 300 K with a Nos\`e-Hoover
 thermostat~\cite{Nose84a,Nose84b,Hoover}.

\subsection{Partially stabilized zirconia}

 The static minimization of the V$_1$ super-cell has already been
 described. Further analysis of the structure reveals that the atomic
 configuration obtained in Section~\ref{isolvac} is metastable,
 corresponding to a saddle point in the potential energy. By perturbing the
 relaxed configuration obtained in Section~\ref{isolvac}, a secondary
 structural modification further lowers the total energy of the cell,
 similar to the tetragonal distortion of the oxygen sublattice in pure
 stoichiometric $c$ zirconia (soft $X_2^-$ mode of vibration).

 According to the energy surface governing the $c-t$
 relationship,~\cite{mdZrO2} the internal tetragonal distortion of the
 oxygen sublattice drives the external tetragonal distortion of the
 unit-cell. As a consequence, the relaxed configuration is expected to be
 further minimized by a non unitary $c/a$ ratio. Indeed, allowing the
 tetragonality of the cell to adjust, the minimum energy is obtained for
 $c/a=1.02$. The corresponding equilibrium structure is shown in
 Figure~\ref{psz}. The anions and cations are represented with light and
 dark circles respectively and the arrows point to the oxygen column
 containing the vacancy, which is behind the first visible oxygen ion. The
 projection on the $x-y$ plane (left) shows the same pattern obtained with
 the static minimization. The NN cations move away from the vacancy and the
 NN anions contract inward. The right panel shows the projection along the
 $x-z$ plane: the tetragonal distortion of the oxygen sublattice is very
 clear. Note the perturbation in the neighborhood of the vacancy arising
 from the superposition of the tetragonal distortion and the
 vacancy-induced radial displacement field.

 The MD simulation of the V$_1$ super-cell shows that this atomic structure
 is stable and does not undergo any further structural modification. The
 atomic configuration obtained by averaging the atomic positions over the
 last 20 ps coincide with the one obtained with the static minimization,
 shown in Figure~\ref{psz}.

 The similarity between the $t^\ast$ and $t$ structures is also evident
 from the Radial Distribution Function (RDF) shown in Figure~\ref{rdf2}. In
 agreement with the experimental
 measurements~\cite{Tuilier87,Morikawa88,Li93a,Li93b,Li93c,Li94a,Li94b,Li94c},
 the splitting of the first peak in the Zr-O RDF, a sign of the oxygen
 sublattice being tetragonally distorted, is also present in the RDF of the
 $t^\ast$ phase. The first peak in the O-O RDF is exactly the same in both
 the tetragonal phases, while the disorder introduced by the vacancy
 slightly modifies the second O-O coordination shell. The Zr-Zr RDF
 indicates that the cation sublattice is nearly $fcc$. Arrows in
 Figure~\ref{rdf2} indicate the available experimental values for 3\% and
 15\% mol Y$_2$O$_3$~\cite{Li93a,Li93b,Li93c,Li94a,Li94b,Li94c}.

\subsection{Fully stabilised zirconia}

 The $c^\ast$ phase was modelled by distributing four vacancies in the
 super-cell. Here we consider two possible high-symmetry vacancy
 configurations. In the first one, the vacancies were distributed in the
 cell by maximizing the defect-defect separation, by setting the four
 vacancies on a $fcc$ lattice, with a defect-defect distance of
 $\sqrt{2}\,a_0$ ($a_0$=9.61 a.u.=5.09 \AA\ is the lattice constant of the
 cubic structure). In the second configuration, the four vacancies were
 aligned along the $\langle 111 \rangle$ direction, with a defect-defect
 minimum distance of $\sqrt{3}/2\,a_0$. The SC-TB model shows that the
 second configuration is more stable than the first one by 0.1 Ry per
 super-cell. It therefore predicts the clustering of vacancies along the
 $\langle 111 \rangle$ direction, as observed
 experimentally~\cite{Goff99,Rossell76,Stubican78} and from
 first-principles calculations.~\cite{Bogicevic01} The results for the most
 stable of the two vacancy configurations, which we refer to as the $V_4$
 super-cell, are described in what follows.

 The static relaxation was started with all the atoms in the fluoritic
 centro-symmetric positions. In this case, only one structural modification
 was observed. The system evolved towards the atomic configuration shown in
 Figure~\ref{fsz}. The local distortions around the vacancies lock the
 $c^\ast$ crystal in a distorted configuration, suppressing the tetragonal
 distortion of the oxygen sublattice. The same structure was then obtained
 by MD simulation of the V$_4$ super-cell. The local atomic environment is
 very different from that of the fluorite structure: 14 cations out of 32
 are 7 fold coordinated and the majority of the oxygen ions are not in
 centro-symmetric positions. The minimum energy is obtained for a perfect
 cubic super-cell ($c/a=1$).

 Despite these distortions, the RDF shown in Figure~\ref{rdf2} suggests
 that the structure of the V$_4$ is, on average, cubic. The first peak of
 the Zr-O component confirms that the oxygen sublattice is not tetragonally
 distorted. The calculated average Zr-O distance compares very well with
 the experimental one in 15\% mol yttria stabilized
 zirconia~\cite{Li93a,Li93b,Li93c,Li94a,Li94b,Li94c}, shown by the arrow.
 In the tetragonal structures, the O-O second coordination shell shows a
 characteristic three-peaked RDF. The corresponding RDF of the $c^\ast$
 phase, shows only a single peak centered at $r=a_0\,\sqrt(2)=6.8$ a.u.,
 which is the second shell O-O distance in the perfect cubic lattice.
 Because of the outward displacement of the large number of Zr atoms NN to
 the vacancies, the first Zr-Zr peak is shifted to smaller $r$, which is
 also experimentally observed (see arrow in Figure~\ref{rdf2}c). Apart from
 this detail, the Zr-Zr RDF shows complete similarity with the other cases
 and therefore the Zr sublattice is $fcc$, even in the V$_4$ supercell.

\section{Discussion}
\label{disc}

 Our results corroborate some of the early experimental observations on
 stabilized zirconia and on other compounds with the general formula
 M$_7$O$_{12}$.~\cite{Rossell76,Stubican78} These reported a strong
 distortion of the oxygen sublattice, with the NN O atoms relaxing towards
 the vacancy by $\approx$ 0.3 \AA\ and the NN Zr relaxing away from the
 vacancy by less than 0.2 \AA.~\cite{Rossell76} Moreover, in the special
 case of the Zr$_3$M$_4$O$_{12}$ compounds, the difference between the
 ordered and disordered fluorite structures was determined by the
 arrangement of the oxygen vacancies,~\cite{Rossell76} which where reported
 to lie along the [111] directions.~\cite{Stubican78}

 When the results described in the previous Sections are considered in the
 context of the $c \leftrightarrow t$ phase transition in pure
 zirconia,~\cite{strcZrO2} they also provide insight on how the structural
 distortions around the oxygen vacancies affect the relative stability of
 the tetragonal and cubic phases. The relationship between the $c$ and $t$
 forms is governed by a temperature-dependent double-well in the potential
 energy. At low temperature, the cubic phase $c$ is structurally unstable
 and it lowers the energy by distorting the oxygen sublattice along the
 $X_2^-$ mode of vibration. The extent of this distortion is measured by
 the order parameter $\delta$. The 0 K double-well for the $c-t$ structures
 is shown in Figure~\ref{dw} by a solid line labelled $t$.

 The static relaxations showed that both the unrelaxed cells V$_1$ and
 V$_4$ are unstable with respect to the short-range isotropic relaxation
 around the vacancies. Moreover, our atomistic simulations suggest that the
 V$_1$ cell is also unstable with respect to the tetragonal distortion. Let
 us now simply superimpose these two displacement fields to see how the
 oxygen vacancies, considered as centers of dilations, modifies the
 energy double-well. In the following, the statically relaxed
 configurations of V$_1$ and V$_4$ which show the radial displacements
 only, are taken as reference systems, to which we apply the tetragonal
 distortion $\delta$. The results are shown in Figure~\ref{dw}.

 Similarly to the $c$ structure, the defective V$_1$ cell is unstable with
 respect to the tetragonal distortion: the atomic relaxations around the
 vacancy partially reduce the height and width of the double well
 (Figure~\ref{dw}).  A static minimization from the minimum of the
 double-well leads to the same structure obtained by averaging over the MD
 configurations (Figure~\ref{psz}), and showing that the finite-temperature
 equilibrium configuration is not just a superposition of the two
 displacement fields.

 On the other hand, the reference V$_4$ cell is stable with respect to the
 tetragonal distortion of the oxygen sublattice and there is no double-well
 in the total energy (Figure~\ref{dw}). The energy profile is highly
 anharmonic, just as the high-temperature one in the free-energy for the
 pure phase~\cite{strcZrO2}. This means that in our model for $c^*$, it is
 possible to shift entire column of oxygen atoms by 5-10 \% of the Zr-O
 interatomic distance at no energy cost, with important implications for
 the mobility of defects and therefore for transport phenomena in
 stabilized zirconia. We know from our previous simulations, and it is not
 surprising, that in a flat well of this form the average structure at
 finite temperature is cubic.

\section{Conclusions}
\label{concl}

 In conclusion, partially and fully stabilized zirconia were modelled by
 introducing respectively one and four vacancies in 96-site cubic
 super-cell (V$_1$ and V$_4$, respectively). Energetics and atomic
 structures of the defective super-cells were calculated with a SC-TB model
 which was parameterized on the properties of pure stoichiometric zirconia.

\begin{figure}
\caption{Equilibrium crystal structure of the V$_1$ supercell. The arrow
 point to the oxygen columns containing the vacancy.}
\label{psz}
\centerline{\psfig{file=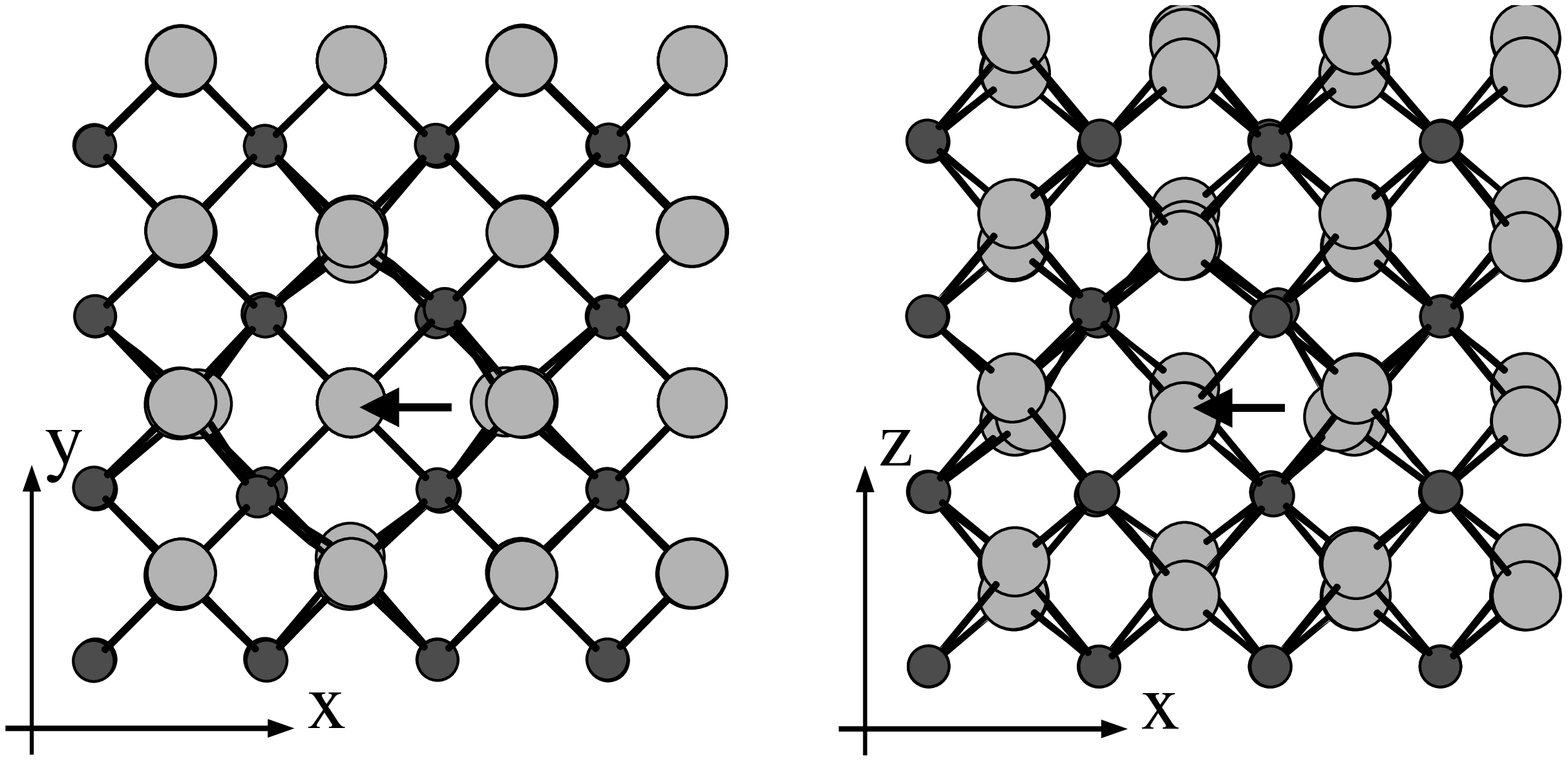,width=9cm,angle=0}} 
\end{figure}

\begin{figure}
\caption{Radial Distribution Function $g(r)$ obtained from the MD
 simulations at 300 K for three supercells: $t$-ZrO$_2$, V$_1$ and
 V$_4$. Arrows represent the experimental values for $t^\ast$- and
 $c^\ast$-ZrO$_2$ (Ref. 13).}
\label{rdf2}
\centerline{\psfig{file=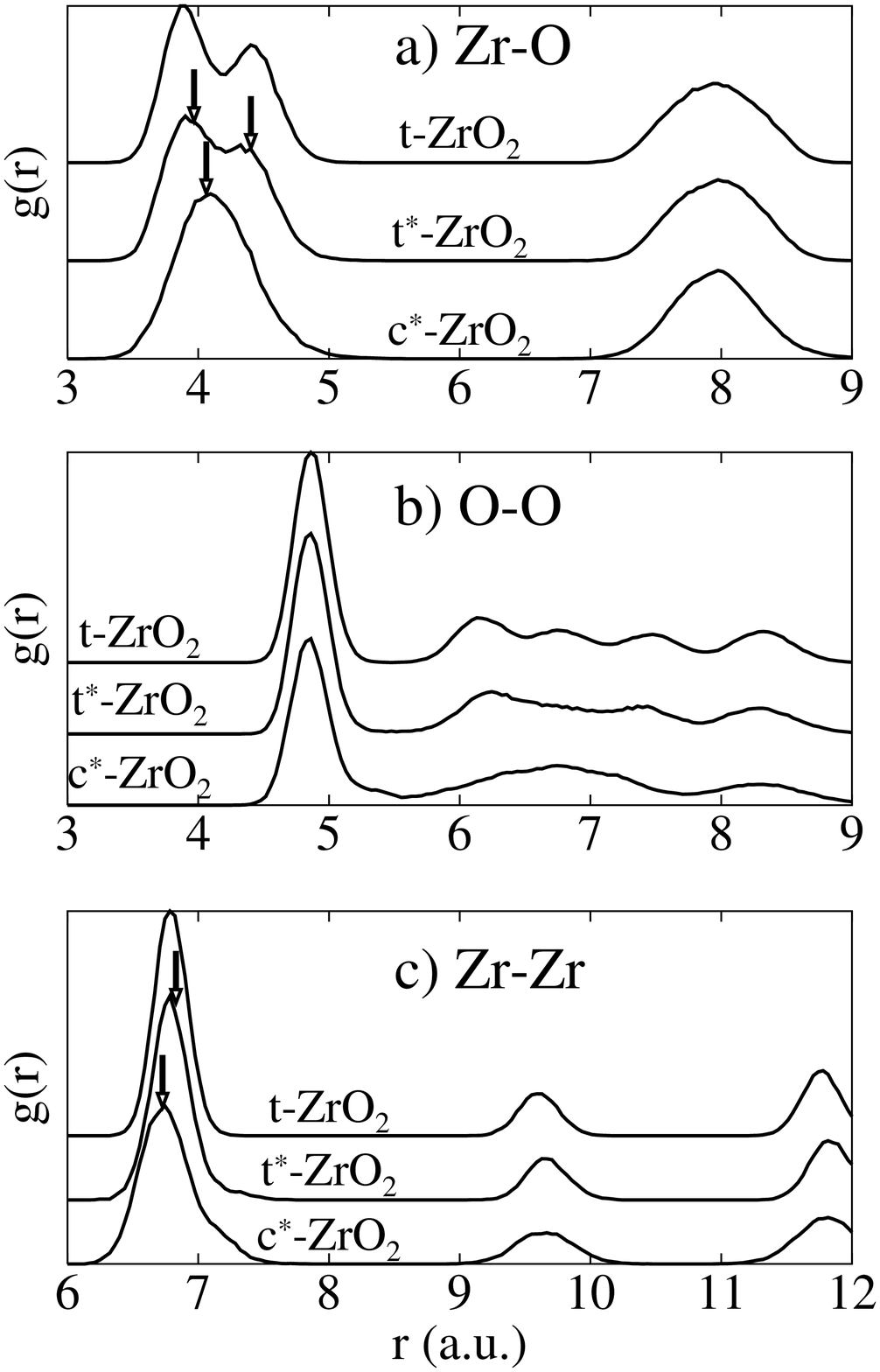,width=9cm,angle=0}} 
\end{figure}

 The analysis of the V$_1$ and V$_4$ equilibrium crystal structures
 suggests a possible explanation of the stabilization mechanism.  When the
 concentration of vacancies is low, as in the V$_1$ super-cell, a
 relatively large volume of crystal is left in the fluorite structure, and
 it undergoes the tetragonal distortion just as the stoichiometric material
 in the $c$-phase would. Because this distortion involves the coordinated
 motion of all the oxygens, also the atoms neighboring the defects are
 dragged along the tetragonal distortion of the oxygen sublattice (see
 Figure~\ref{psz}).

\begin{figure}
\caption{Equilibrium crystal structure of the V$_4$ supercell. The arrows
 point to the oxygen columns containing the vacancies.}
\label{fsz}
\centerline{\psfig{file=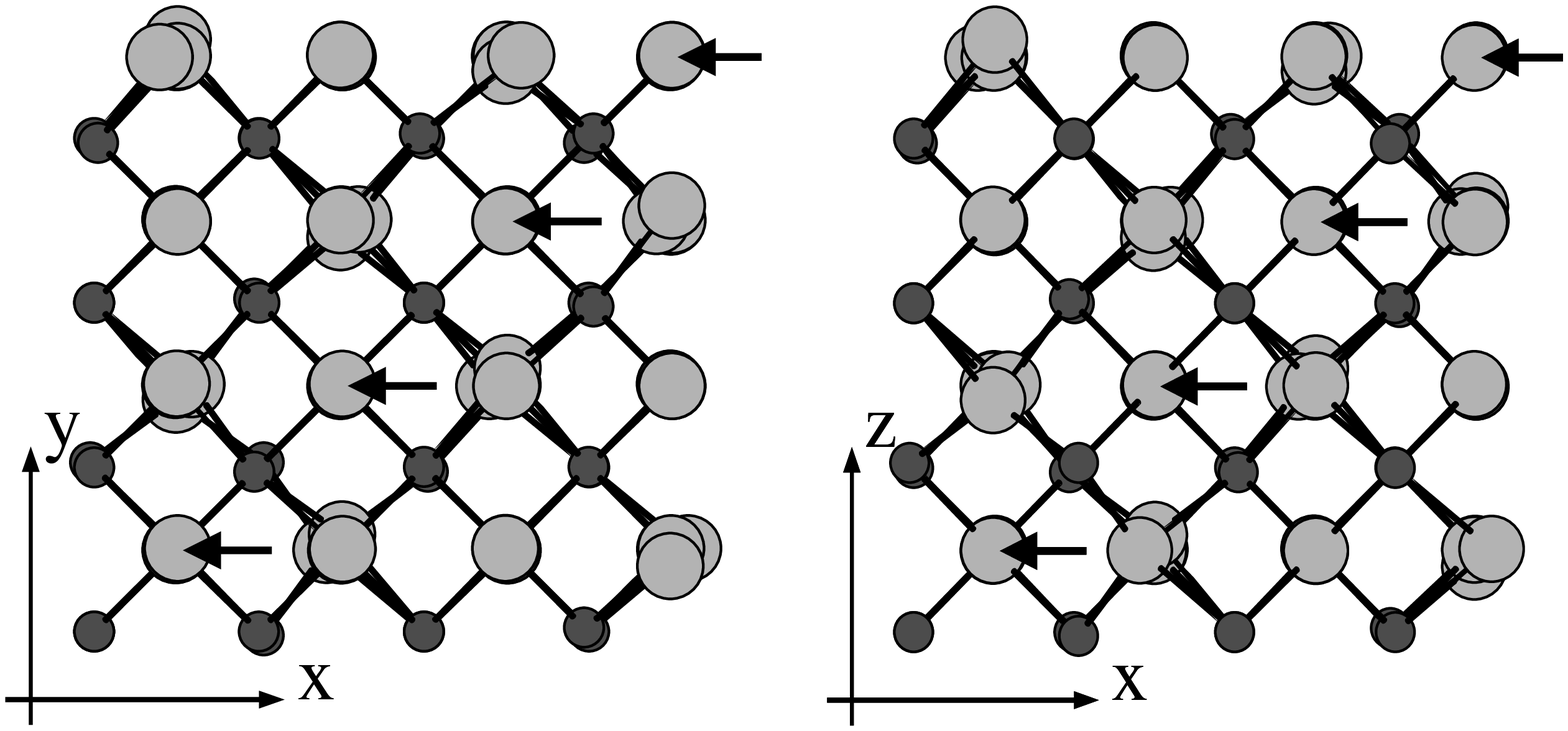,width=9cm,angle=0}} 
\end{figure}

\begin{figure}
\caption{Total energy vs. tetragonal distortion $\delta$ of the oxygen
sublattice for three 96-site cells: $t$ pure stoichiometric tetragonal 
structure; $t^*$  partially stabilized zirconia ($V_1$ super-cell); $c^*$
fully stabilized zirconia ($V_4$ super-cell).}
\label{dw}
\centerline{\psfig{file=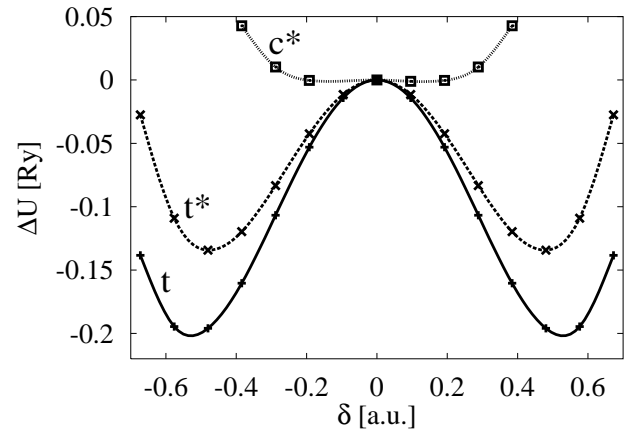,width=9cm,angle=-90}} 
\end{figure}

 When the concentration of defects is high, as in the V$_4$ super-cell,
 there is virtually no undistorted cubic region in the statically relaxed
 super-cell; every oxygen atom is either itself a neighbor of a vacant
 site, or at least four of its six neighboring oxygen atoms are. Thus there
 is virtually no region in which the local atomic environment is close to
 fluorite, which could undergo the tetragonal distortion (see
 Figure~\ref{fsz}), and the radial distortions around the vacancies
 dominate the finite-temperature equilibrium configuration.  The resulting
 atomic structure is cubic only by averaging over relatively large number
 of atoms ($\approx $ 100), but the short-range atomic structure does not
 have cubic symmetry.

 These results show that the stabilization of the $t^\ast$ and $c^\ast$
 structures may be achieved in theory by doping zirconia crystals with
 oxygen vacancies only, and support the idea that the electronic and
 structural properties of stabilized zirconia are controlled by the
 structural disorder around the oxygen vacancies, rather than by the
 substitutional dopant cations.



\newpage
\centerline{FIGURE CAPTIONS}

FIG.1.{Crystal relaxation of the atoms neighbouring an isolated vacancy
 in a fluorite lattice. Key: Zr black, O dark grey, vacancy light grey.}

FIG.2.{Total density of states for the relaxed and unrelaxed
 configurations of the $V_1$ supercell, calculated with the SC-TB. The 0
 of energy is the highest occupied electron state. The arrow points to the
 energy levels of the Zr atoms NN to the vacancy, split from the $d$
 bands of the ``bulk'' Zr atoms.}

FIG.3.{Equilibrium crystal structure of the V$_1$ supercell. The arrow
 point to the oxygen columns containing the vacancy.}

FIG.4.{Equilibrium crystal structure of the V$_4$ supercell. The arrows
 point to the oxygen columns containing the vacancies.}

FIG.5.{Radial Distribution Function $g(r)$ obtained from the MD
 simulations at 300 K for three supercells: $t$-ZrO$_2$, V$_1$ and
 V$_4$. Arrows represent the experimental values for $t^\ast$- and
 $c^\ast$-ZrO$_2$ (Ref. 13).}

FIG.6.{Total energy vs. tetragonal distortion $\delta$ of the oxygen
sublattice for three 96-site cells: $t$ pure stoichiometric tetragonal 
structure; $t^*$  partially stabilized zirconia ($V_1$ super-cell); $c^*$
fully stabilized zirconia ($V_4$ super-cell).}

\begin{table}
\caption{Atomic displacement (in \AA) predicted by the SC-TB model
 compared to ab initio results.$^1$}
\label{tab}
\begin{center}
 \begin{tabular}{ll|cccc}
    &    & O  NN & Zr NN & O  NNN & Zr NNN \\ \hline
DFT   & Ref.~\onlinecite{Stapper99} & 0.24 & 0.14 & 0.04 & 0.04 \\
SC-TB & this work                   & 0.27 & 0.18 & 0.03 & 0.03 \\
  \end{tabular}
\end{center}
\end{table}
$^1${The displacements are along the
 directions shown by the arrows in Figure 1. NN and NNN mean Nearest
 Neighbour and Next Nearest Neighbour to the defect.}

\newpage

\end{document}